\documentclass[12pt]{iopart}

\usepackage[hmargin=1.5cm,vmargin=2cm]{geometry}
\usepackage{graphicx}
\usepackage{amssymb}
\usepackage{bm}
\usepackage[table]{xcolor}
\usepackage[normalem]{ulem} 
\usepackage[colorlinks=true,linkcolor=blue,citecolor=blue,urlcolor=blue]{hyperref}

\begin{document}

\newcommand{\op}[1]{{\bm{#1}}}
\newcommand{\opcal}[1]{\boldsymbol{\mathcal{#1}}}
\newcommand{\new}[1]{\textcolor[rgb]{0,0,1}{#1}}
\newcommand{\old}[1]{\textcolor[rgb]{0.2,0.2,0.2}{\sout{#1}}}
\newcommand{\nota}[1]{\textcolor[rgb]{1,0,0}{{#1}}}

\newcommand{\orcid}[1]{\href{https://orcid.org/#1}{\textcolor[rgb]{0.651,0.808,0.224}{\includegraphics[width=10pt]{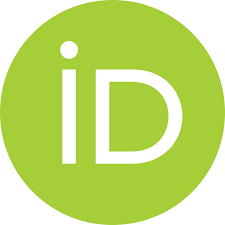}}}}

\title[Dynamic violation of Bell's inequalities\dots]{Dynamic violation of Bell's inequalities in the angular momentum representation}

\author{J.A. L\'opez--Sald\'ivar{$^{1,2}$} \orcid{0000-0001-5489-698X}, O. Casta\~nos{$^{3}$} \orcid{0000-0003-2514-3764}, S. Cordero{$^3$} \orcid{0000-0003-3594-3992},  R. {L\'opez--Pe\~na}{$^3$} \orcid{0000-0003-2214-5991}, and E. {Nahmad--Achar}{$^3$} \orcid{0000-0003-3707-8293}}

\address{
$^1$
Russian Quantum Center, Skolkovo, Moscow 143025, Russia
}
\address{
$^2$ National University of Science and Technology “MISIS”, Moscow 119049, Russia
}
\address{%
{$^3$} Instituto de Ciencias Nucleares, Universidad Nacional Aut\'onoma de M\'exico, Apartado Postal 70-543, 04510  Mexico City, Mexico}

\ead{\mailto{julio.lopez.8303@gmail.com}, \mailto{ocasta@nucleares.unam.mx}, \mailto{sergio.cordero@nucleares.unam.mx}, \mailto{lopez@nucleares.unam.mx},  \mailto{nahmad@nucleares.unam.mx}}


\begin{abstract}
A parametrization of density matrices of $d$ dimensions in terms of the raising $\op{J}_+$ and lowering $\op{J}_-$ angular momentum operators is established together with an implicit connection with the generalized Bloch-GellMann parameters. A general expression for the density matrix of the composite system of angular momenta $j_1$ and $j_2$ is obtained.  In this matrix representation violations of the Bell-Clauser-Horne-Shimony-Holt inequalities are established for the $X$-states of a qubit-qubit, pure and mixed, composite system, as well as for a qubit-qutrit density matrix. In both cases maximal violation of the Bell inequalities can be reached, i.e., the Cirel'son limit. A correlation between the entanglement measure and a strong violation of the Bell factor is also given. For the qubit-qutrit composite system a time-dependent convex combination of the density matrix of the eigenstates of a two-particle Hamiltonian system is used to determine periodic maximal violations of the Bell's inequality.
\end{abstract}

\maketitle

\section{Introduction}

The Einstein-Podolsky-Rosen paradox argues that quantum mechanics is not a complete theory and that it should be supplemented with additional variables called {\it hidden variables}~\cite{epr35}. However, Bell showed that measurements carried out on two entangled states spatially separated cannot be reproduced using local hidden variables~\cite{bell64, bell66}. One must then consider experimental settings of the Bohm-Aharonov type, where the state is allowed to change during the flight of the particles~\cite{bohm57}. A generalization of Bell's inequalities was given by the works of Clauser-Horne-Shimony-Holt (CHSH)~\cite{clauser78}, which was more suitable for direct experiments. The experiments realized by Freedman and Clauser, and by Aspect et al., show agreement with the quantum mechanical predictions, yielding a strong violation of Bell's inequalities and the rejection of realistic local theories, implying that nature cannot be described by a local realistic model~\cite{freedman72, aspect82, lo81,aspect76}. 

The Bell-CHSH inequalities can be used as a criterion for a classical description, that is, to quantify the non-locality present in the correlations obtained in experiments by considering local hidden variables. This statement has a deep meaning by considering that quantum mechanics is universal and that the classical reality can be obtained from the quantum dynamics. The mentioned inequalities are bounded by the value of $2$ in local hidden-variable models. In particular, for the entangled quantum bipartite systems Bell's inequality has a limit of $2 \sqrt{2}$.~\cite{tsirelson92}. A demonstration of this upper bound is given in~\cite{tsirelson80, buhrman2005}. 

The Bell inequalities have been very useful in quantum information theory because they allow to separate the classical correlations from the quantum correlations, and there are many quantum information protocols that make use of Bell states or maximally entangled two-particle states, in particular qubit-qubit, qubit-qutrit and in general qubit-qudit systems~\cite{leslie2019, mendez2023, bernal2024}. 
Thus one finds many studies about the violation of the Bell-CHSH inequality in bipartite systems described by pure states (cf. e.g.,~\cite{ sorella2024} where they find that it depends strongly on the way the observables for Alice and Bob are selected).  For a system of two spins of angular momentum $j$ the entanglement and the maximal violation of the CHSH inequality have been studied, in agreement with previous work by Gisin-Peres~\cite{gisin92a, peruzzo2023}. The contribution of Gisin-Peres shows that the maximum violation of Bell's inequality occurs for arbitrarily large spin particles, implying that the classical properties do not always emerge for large quantum systems.  A finite violation of $24$\% of the times is obtained for arbitrarily large $j$ spin particles prepared in a singlet state; additionally, an optical realization of the singlet $j$-state was established by means of two traveling-wave modes of a quantized field~\cite{peres92, gerry2005}. 

A composite quantum state is classically correlated if it can be written as a convex combination of product states; when this is not possible, the state has been called an EPR correlated state. Then, according to Bell's theorem, a classically correlated state can be described by a local hidden-variable theories and satisfies the Bell inequality. On the other hand, if a state satisfies the Bell inequality it is not always true that it is a classically correlated state.~\cite{werner89}. 

In the first decade of this century the so-called {\it X states} have been studied in quantum information theory because the Bell entangled states and the Werner states are subsets of them. The X states have $7$ free  parameters whose invariance algebra is $\mathfrak{su}(2) \times \mathfrak{su}(2) \times \mathfrak{u}(1)$. It is a subalgebra of $\mathfrak{su}(4)$, the complete symmetry of a two-qubit system~\cite{rau2009}.  Notice that there are $15$ different ways to generate the subalgebra $\mathfrak{su}(2) \times \mathfrak{su}(2) \times \mathfrak{u}(1)$ according to the selection of the commuting element (the generator of $\mathfrak{u}(1)$)~\cite{kelleher2021}. Each one of them constitutes a Fano plane in the projective geometry PG(3,2). In particular the set of traceless operators,
\begin{eqnarray}
\{&&\op{X}_1 = \op{\sigma}_z \otimes \op{\sigma}_z, \op{X}_2= \op{\sigma}_y \otimes \op{\sigma}_x , \op{X}_3= \op{I}_2 \otimes \op{\sigma}_z, \op{X}_4= -\op{\sigma}_y \otimes \op{\sigma}_y,\nonumber\\ 
&&\op{X}_5 = \op{\sigma}_x \otimes \op{\sigma}_y , \op{X}_6= \op{\sigma}_z \otimes \op{I}_2, \op{X}_7= \op{\sigma}_x \otimes \op{\sigma}_y \} \, ,
\label{eqX}
\end{eqnarray}
has been used in the context of quantum logic gates~\cite{rau2000}. 

A review of the parametrization of finite density matrices can be found in~\cite{petruccione2012}.  There, they discuss the following parametrizations: Bloch-vector, the polarization operator basis, the coset, and the Jarlskog, together with density matrices associated to composite systems. Another relevant review for the possible applications of the present contribution is associated to the quantum information theory (cf.~\cite{horodecki21} and references therein.

In this contribution a parametrization of the finite density matrices is considered in terms of the expectation values of the angular momentum operators. First we construct the density matrix for a qubit using the linear power of the generators of the angular momentum algebra; for the qutrit one needs eight independent operators, which can be obtained by considering up to quartic products of the normalized raising and lowering angular momentum operators. For the ququart one has to consider up to the sixth (combined) power of these operators. The results can be generalized for the case of a qudit, that is, a Hilbert space of $d$ dimensions with $d=2 j +1$, which includes the expectation values of the normalized raising and lowering operators up to a power equal to $4 j$. This representation of the density matrix has some advantages over the standard parametrizations. One of them is the possibility to write the state in terms of expectation values of observables (which can be done by expressing the operators $\op{J}_\pm$ in terms of $\op{J}_x$ and $\op{J}_-$). Another advantage is in the study of the entanglement.

The general parametrization of a composite system consisting of the tensorial product of two qudits, that is associated to the unitary algebras $\mathfrak{su}(2j_1+1)$ and $\mathfrak{su}(2 j_2+1)$, is here established.  The separability properties are given together with the corresponding reduced density matrix elements, then the partial transpose criterion is used to determine the quantum correlations of the composite system. Since pair-wise entanglement of quantum systems promises to be of benefit to secure quantum information transmission and to quantum computing, and this entanglement is measured via the linear and von Neumann entropies calculated through reduced density matrices, our parametrization allows us to study not only the phase diagrams of interacting systems, but the quantum correlations between subsystems in a bipartite set-up.

An interesting example is the evaluation of the Clauser-Horne-Shimony-Holt (CHSH) inequality~\cite{clauser69}, which we do in Section \ref{sec3} for a two-qubit system and for a qubit-qutrit system, for which values reaching the Cirel'son bound are obtained.

\section{Density matrix parametrization using angular momentum mean values}
The density matrix of a general quantum system can be parametrized using the mean values of monomial products of the operators $\op{J}_\kappa (\kappa=x,y,z)$. To show this we present the cases for $j=1/2$, $j=1$, and $j=3/2$, and with the knowledge from these we generalize to any qudit system.

It is well known that the density matrix of a qubit can be written in terms of the expectation values of the Pauli matrices $\op{\sigma}_k$ as
\begin{equation}
\op{\rho}=\frac{1}{2}\left(\begin{array}{cc}
1+\lambda_3 & \lambda_1-i\lambda_2 \\
\lambda_1+i \lambda_2 & 1-\lambda_3
\end{array}\right).
\end{equation}
where $\lambda_k ={\rm Tr}(\op \rho \, \op \sigma_k ) = \langle\op{\sigma}_k\rangle$. The polarization vector, used in the field of quantum optics, is given by $\vec{P} = 2\langle \vec{\op{J}} \rangle$ with $\vec{\op{J}}=\vec{\op{\sigma}}/2$ (here, and hereafter we set $\hbar=1$).

This density matrix can describe pure and mixed qubit states, where the probability of finding the qubit in the state $\vert 1 \rangle$ ($\vert 0 \rangle$) is given by $p_1 = (1/2)(1+\lambda_3)$ ($p_0 = (1/2)(1-\lambda_3)$). The non-diagonal terms represent the coherences of the transition amplitudes. The purity of the state is given by $\Tr(\op{\rho}^2) = (1/2)(1 + \vert\vec{\lambda}\vert^2)$, i.e., a pure state has $\vert\vec{\lambda}\vert=1$ and the most mixed state has $\vert\vec{\lambda}\vert=0$.

This description has been extended to describe a qudit system in terms of the generalized Gell-Mann matrices
\begin{equation}
\op{\rho} = \frac{1}{d} \op{I}_d + \frac{1}{2}\sum_{k=1}^{d^2-1} \lambda_k \op{\Lambda}_k
\end{equation}
which satisfy $\Tr({\op{\Lambda}_k})=0,\ \Tr({\op{\Lambda}_j \op{\Lambda}_k})=2\delta_{jk}$, and where $\lambda_k = \langle\op{\Lambda}_k\rangle$.

We may parametrize the density matrix in terms of the angular momentum operators, i.e., the qutrit represented by $J=1,\ (d=3)$, the ququart by $J=3/2,\ (d=4)$, and so on. For the qubit, one has enough independent operators by considering the expectation values of the angular momentum. Rewriting the density matrix in terms of the operators $\op{J}_\pm =\op{J}_x \pm i \op{J}_y$, and using $\op{J}_+ \op{J}_-=\frac{1}{2}\op{I} + \op{J}_z$ and $\op{J}_- \op{J}_+= \frac{1}{2}\op{I}-\op{J}_z$ we get
\begin{equation}
\op{\rho}=\left(\begin{array}{cc}
\langle \op{J}_+ \op{J}_-\rangle & \langle \op{J}_-\rangle\\
\langle \op{J}_+ \rangle & \langle \op{J}_- \op{J}_+ \rangle
\end{array}
\right).
\end{equation}
In this case, the generators of the $\mathfrak{su}(2)$ algebra (the Pauli matrices $\op{\sigma}_{1,2,3}$) are proportional to the angular momentum operators as $\op{J}_k=\op{\sigma}_k/2$.

For the qutrit, on the other hand, one needs $8$ independent operators, for which we take
\begin{equation}
\mathcal{S}_3=\{\op{J}_+^2, \op{J}_-^2, \op{J}_+ \op{J}_-^2, \op{J}_+^2 \op{J}_-, \op{J}_-^2 \op{J}_+, \op{J}_- \op{J}_+^2, \op{J}_+^2 \op{J}_-^2, \op{J}_- \op{J}_+^2 \op{J}_-, \op{J}_-^2 \op{J}_+^2 \}.
\end{equation}
in terms of which the density operator takes the form
\begin{equation}
\op{\rho} = \frac{1}{4} \left(
\begin{array}{ccc}
 \langle \op{J}_+^2 \op{J}_-^2 \rangle & \sqrt{2} \langle \op{J}_+ \op{J}_-^2\rangle & 2 \langle \op{J}_-^2 \rangle \\[1mm]
 \sqrt{2} \langle \op{J}_+^2 \op{J}_- \rangle & \langle \op{J}_- \op{J}_+^2 \op{J}_- \rangle& \sqrt{2} \langle \op{J}_-^2 \op{J}_+ \rangle \\[1mm]
 2 \langle \op{J}_+^2 \rangle & \sqrt{2} \langle \op{J}_- \op{J}_+^2 \rangle & \langle \op{J}_-^2 \op{J}_+^2 \rangle
\end{array}
\right) =
\left(
\begin{array}{ccc}
 \langle \op{\mathcal{J}}_+^2 \op{\mathcal{J}}_-^2 \rangle & \langle \op{\mathcal{J}}_+ \op{\mathcal{J}}_-^2\rangle & \langle \op{\mathcal{J}}_-^2 \rangle \\[1mm]
 \langle \op{\mathcal{J}}_+^2 \op{\mathcal{J}}_- \rangle & \langle \op{\mathcal{J}}_- \op{\mathcal{J}}_+^2 \op{\mathcal{J}}_- \rangle& \langle \op{\mathcal{J}}_-^2 \op{\mathcal{J}}_+ \rangle \\[1mm]
 \langle \op{\mathcal{J}}_+^2 \rangle & \langle \op{\mathcal{J}}_- \op{\mathcal{J}}_+^2 \rangle & \langle \op{\mathcal{J}}_-^2 \op{\mathcal{J}}_+^2 \rangle
\end{array}
\right)\ ,
\end{equation}
where we have defined $\op{\mathcal{J}}_\pm :=\op{J}_\pm/\sqrt{2}$.
This may also be written using the diagonal operators
\begin{eqnarray}
\op{F}(\op{J}_z) &:=& \op{J}^2 - \op{J}_z^2 + \op{J}_z= \op{J}_{+}\op{J}_{-}\ ,  \nonumber\\
\op{G}(\op{J}_z) &:=& \op{J}^2 - \op{J}_z^2 - \op{J}_z = \op{J}_{-}\op{J}_{+}\ ,
\end{eqnarray}
 and powers of  the ladder operators, as shown in the Appendix B.

In the ququart case we have the 15 generalized Gell-Mann matrices $\op{\lambda}_{1,\ldots,15}$. Inspired by the results for the qutrit case we propose the the following set of angular momentum operators
\begin{eqnarray}
\mathcal{S}_4&=&\{\op{J}_+^2 \op{J}_-^3,\op{J}_+ \op{J}_-^3,\op{J}_-^3,\op{J}_+\op{J}_-^3\op{J}_+,\op{J}_-^3\op{J}_+,\op{J}_-^3\op{J}_+^2,\op{J}_+^3\op{J}_-^3,\op{J}_+^2\op{J}_-^3\op{J}_+, \nonumber \\[1mm]
&& \op{J}_+\op{J}_-^3\op{J}_+^2,\op{J}_-^3\op{J}_+^3,\op{J}_+^3\op{J}_-^2,\op{J}_+^3\op{J}_-,
\op{J}_+^3, \op{J}_-\op{J}_+^3\op{J}_-,\op{J}_-\op{J}_+^3,\op{J}_-^2\op{J}_+^3 \}.
\end{eqnarray}
In this case the density matrix can be written as
\begin{equation}
\op{\rho} = \left(\begin{array}{cccc}
\langle \op{\mathcal{J}}_+^3 \op{\mathcal{J}}_-^3\rangle &  \langle \op{\mathcal{J}}_+^2 \op{\mathcal{J}}_-^3\rangle & \langle \op{\mathcal{J}}_+ \op{\mathcal{J}}_-^3 \rangle & \langle \op{\mathcal{J}}_-^3\rangle\\[1mm]
\langle \op{\mathcal{J}}_+^3 \op{\mathcal{J}}_-^2\rangle & \langle\op{\mathcal{J}}_+^2 \op{\mathcal{J}}_-^3 \op{\mathcal{J}}_+\rangle & \langle \op{\mathcal{J}}_+ \op{\mathcal{J}}_-^3 \op{\mathcal{J}}_+\rangle & \langle \op{\mathcal{J}}_-^3 \op{\mathcal{J}}_+ \rangle \\[1mm]
\langle \op{\mathcal{J}}_+^3 \op{\mathcal{J}}_-\rangle & \langle \op{\mathcal{J}}_- \op{\mathcal{J}}_+^3 \op{\mathcal{J}}_-\rangle & \langle \op{\mathcal{J}}_+ \op{\mathcal{J}}_-^3 \op{\mathcal{J}}_+^2 \rangle & \langle \op{\mathcal{J}}_-^3 \op{\mathcal{J}}_+^2\rangle \\[1mm]
\langle \op{\mathcal{J}}_+^3\rangle & \langle \op{\mathcal{J}}_- \op{\mathcal{J}}_+^3\rangle & \langle \op{\mathcal{J}}_-^2 \op{\mathcal{J}}_+^3\rangle & \langle \op{\mathcal{J}}_-^3 \op{\mathcal{J}}_+^3\rangle
\end{array} \right)\, ,
\end{equation}
where the operators $\op{\mathcal{J}}^r_\pm=\sqrt{\frac{(3-r)!}{3! \, r!}} \op{J}_\pm^r$.

Finally, to obtain a general expression for a qudit system, we make use of the angular momentum states for an arbitrary algebra $\mathfrak{su}(d)$: $\vert j,m \rangle$, with $m=-j,-j+1,\ldots, j-1,j$. These states are related to the canonical basis $\vert k \rangle$ with $k=1,2,\ldots, 2j+1$ in the following way
\begin{equation}
\vert j,j\rangle=\left(\begin{array}{cccc}
1 \\
0\\
\vdots \\
0
\end{array}
\right), \quad \vert j,j-1\rangle=
\left(\begin{array}{ccccc}
0 \\
1\\
0\\
\vdots \\
0
\end{array}
\right), \cdots,
\vert j,-j\rangle=\left(\begin{array}{cccc}
0 \\
\vdots \\
0 \\
1
\end{array}
\right).
\end{equation}
In other words, the $k$-th element of the canonical basis $\vert k\rangle$ is related to the angular momentum state as $\vert k \rangle=\vert j, j-k+1 \rangle$. From this association, the projector $\vert k \rangle \langle l \vert$ can be written as
\begin{equation}
\vert k \rangle \langle l \vert= \vert j,j-k+1 \rangle \langle j, j-l+1 \vert.
\end{equation}
Note that a state with arbitrary projection $m$ can be obtained by the recursive application of the ladder operator $\op{J}_-$ to the maximum weight state $\vert j, j \rangle$, or by applying the operator $\op{J}_+$ recursively to the minimum weight state $\vert j, -j \rangle$
\begin{equation}
\vert j, j-r\rangle=\sqrt{\frac{(2j-r)!}{r! \, (2j)!}}\, \op{J}_-^r \, \vert j, j \rangle \equiv \op{\mathcal{J}}_{-}^r\,\vert j,j\rangle, \quad \vert j,-j+s \rangle= \sqrt{\frac{(2j-s)!}{s! \, (2j)!}} \, \op{J}_+^s \,\vert j,-j \rangle \equiv \op{\mathcal{J}}_{+}^s\,\vert j,-j\rangle
\label{ladder}
\end{equation}
where in the cases $r=0$ or $s=0$ we make the substitution $\op{J}_\pm^0 \rightarrow \op{I}$.

There are various ways to obtain the projector $\vert k \rangle \langle l \vert =: \op{A}_{kl}$ which constitute the generators of an algebra $\mathfrak{u}(d)$ with $d=2j+1$; here we use (see Appendix A for other forms)
\begin{equation}
\op{A}_{kl} = \op{\mathcal{J}}_+^{2j-k+1}\vert j,-j \rangle \langle j,j \vert \op{\mathcal{J}}_+^{l-1},
\end{equation}
where the projector $\vert j,j \rangle \langle j,-j \vert$ can be obtained from Eq.~(\ref{ladder}).

As
\begin{equation}
\op{A}_{d\,1} = \vert j,\,-j \rangle \langle j,\,j\vert = \frac{\op{J}_{-}^{2j}}{(2j)!}
\end{equation}
we may write, in general,
\begin{equation}
\op{A}_{kl} = \op{\mathcal{J}}_{+}^{2j-k+1} \op{\mathcal{J}}_{-}^{2j} \op{\mathcal{J}}_{+}^{l-1} \,.
\label{projectors}
\end{equation}

From the projectors we can construct the generators for the $\mathfrak{su}(d)$ algebra $\op{\Lambda}_r$ where $r=1,2,\ldots,d^2-1$,  and which can be grouped into symmetrical, anti-symmetrical, and diagonal operators. We have $\frac{d(d-1)}{2}$ symmetrical ($\op{\Lambda}_r^{(s)}$) and $\frac{d(d-1)}{2}$ antisymmetrical ($\op{\Lambda}_r^{(a)}$) operators which are defined as
\begin{eqnarray}
\op{\Lambda}_{r}^{(s)}=\op{A}_{kl} + \op{A}_{lk} =\op{\mathcal J}_+^{2j-k+1} \op{\mathcal J}_-^{2j} \op{\mathcal J}_+^{l-1} +\op{\mathcal J}_-^{l-1} \op{\mathcal J}_+^{2j} \op{\mathcal J}_-^{2j-k+1}, \nonumber \\
\op{\Lambda}_{r}^{(a)}=i(\op{A}_{lk}-\op{A}_{kl}) =i (\op{\mathcal J}_-^{l-1} \op{\mathcal J}_+^{2j} \op{\mathcal J}_-^{2j-k+1}-\op{\mathcal J}_+^{2j-k+1} \op{\mathcal J}_-^{2j} \op{\mathcal J}_+^{l-1}),
\label{l_symm_asymm}
\end{eqnarray}
with $k,l=1,2,\ldots,d$; $r=1,2,\ldots,\frac{d(d-1)}{2}$.

Similarly, there are $d-1$ diagonal operators, defined as
\begin{eqnarray}
\op{\Lambda}_r^{(diag)} &=& \sqrt{\frac{2}{r(r+1)}}\left(\sum_{k=1}^{r} \op{A}_{kk} - r \op{A}_{r+1\,r+1}\right) \nonumber \\
&=& \sqrt{\frac{2}{r(r+1)}}\left( \sum_{k=1}^{r} (\op{\mathcal J}_+^{2j-k+1} \op{\mathcal J}_-^{2j} \op{\mathcal J}_+^{k-1})- r \op{\mathcal J}_+^{2j-r} \op{\mathcal J}_-^{2j} \op{\mathcal J}_+^{r} \right),
\label{l_diag}
\end{eqnarray}
with $r=1,\ldots,d-1$. The Gell--Mann generators written as in Eqs.~(\ref{l_symm_asymm}) and~(\ref{l_diag}) allow us to represent any density matrix, Hermitian operator, or unitary operator, in terms of angular momentum operators and their mean values, by using the standard definitions
\begin{eqnarray}
\op{O}= \frac{{\rm Tr}(\op{O})}{d} \op{I}+\frac{1}{2} \sum_{k=1}^{d^2-1} {\rm Tr}(\op{O} \op{\Lambda}_k) \op{\Lambda}_k,\quad \op{U}=\exp \left(i \sum_{k=1}^{d^2-1} \theta_k \op{\Lambda}_k \right)\,,
\end{eqnarray}
where
\begin{equation*}
\op{\Lambda}_k = \left\{ 
\begin{array}{ll}
\op{\Lambda}_k^{(s)} & 1\leq k \leq \frac{d(d-1)}{2} \\
\op{\Lambda}_{k-\frac{d(d-1)}{2}}^{(a)} & \frac{d(d-1)}{2} < k \leq d(d-1) \\
\op{\Lambda}_{k- d(d-1)}^{(diag)} & d(d-1) < k \leq d^2-1 \ .
\end{array}
\right.
\end{equation*}
From the previous results for the qubit, qutrit, and ququart, the general density matrix for a qudit system takes the form
\begin{equation}
\resizebox{0.93\hsize}{!}{$
\op{\rho}=\left(
\begin{array}{cccccc}
\langle \op{\mathcal J}_+^{2j} \op{\mathcal J}_-^{2j}\rangle & \langle  \op{\mathcal J}_+^{2j-1} \op{\mathcal J}_-^{2j}\rangle & \langle\op{\mathcal J}_+^{2j-2} \op{\mathcal J}_-^{2j} \rangle & \cdots & \langle \op{\mathcal J}_+ \op{\mathcal J}_-^{2j} \rangle & \langle \op{\mathcal J}_-^{2j} \rangle \\
 & \langle \op{\mathcal J}_+^{2j-1} \op{\mathcal J}_-^{2j} \op{\mathcal J}_+ \rangle & \langle \op{\mathcal J}_+^{2j-2} \op{\mathcal J}_-^{2j} \op{\mathcal J}_+ \rangle & \cdots & \langle\op{\mathcal J}_+ \op{\mathcal J}_-^{2j} \op{\mathcal J}_+\rangle & \langle\op{\mathcal J}_-^{2j} \op{\mathcal J}_+ \rangle\\
 & & \ddots & \vdots & \vdots & \vdots\\
 & & & \ddots & \vdots & \vdots \\
 & & & & \langle \op{\mathcal J}_+ \op{\mathcal J}_-^{2j} \op{\mathcal J}_+^{2j-1} \rangle & \langle \op{\mathcal J}_-^{2j} \op{\mathcal J}_+^{2j-1} \rangle \\
 & & & & & \langle \op{\mathcal J}_-^{2j} \op{\mathcal J}_+^{2j} \rangle
\end{array}
\right)\,,
$}
\end{equation}
where $\rho_{kl}$ is explicitly written as
\begin{equation}
\rho_{kl} = \Tr(\op{\rho}\,\op{A}_{lk}) = \left\langle \op{\mathcal J}_+^{2j-l+1} \op{\mathcal J}_-^{2j} \op{\mathcal J}_+^{k-1} \right\rangle.
\end{equation}
We note that all the entries of the density matrix can be obtained using the following procedure: first we set $\rho_{11} = \langle \op{\mathcal J}_+^{2j} \op{\mathcal J}_-^{2j}\rangle$, and in the successive columns we diminish the power of $\op{\mathcal J}_+$ by one, while for successive rows, we multiply on the right by the operator $\op{\mathcal J}_+$ before obtaining the expectation value.

This representation of the density matrix has some advantages over the standard parametrizations. One of them is the possibility to write the state in terms of expectation values of observables (which can be done by expressing the operators $\op{\mathcal J}_\pm$ in terms of $\op{\mathcal J}_x$ and $\op{\mathcal J}_y$). Another advantage is in the study of entanglement, as we do in what follows.

\section{Bipartite entanglement properties.}
\label{sec3}

\noindent We will consider two interacting subsystems associated to the unitary algebras $\mathfrak{su}(2j_1+1)$ and $\mathfrak{su}(2 j_2+1)$ respectively. The generators $\op{A}_{k_1 l_1} \otimes \op{A}_{k_2 l_2}$ can be written in terms of the corresponding angular momentum generators (cf. eq.(\ref{projectors}))
\begin{equation}
\op{A}_{k_1\,l_1} \otimes \op{A}_{k_2\,l_2} = (\op{\mathcal J}_+^{2j_1-k_1+1} \op{\mathcal J}_-^{2j_1} \op{\mathcal J}_+^{l_1-1}) \otimes (\op{\mathcal J}_+^{2j_2-k_2+1} \op{\mathcal J}_-^{2j_2} \op{\mathcal J}_+^{l_2-1}),
\end{equation}
where $k_1,l_1=1,2,\ldots,d_1$, $k_2,l_2=1,2,\ldots,d_2$, with $d_1=2j_1+1$ and $d_2=2j_2+1$ representing the dimensions of the two subsystems. This allows us to write the bipartite density matrix elements as the following expectation values
\begin{equation}
\rho_{\tilde{k}\tilde{l}} = \Tr(\op{\rho}\,\op{A}_{l_1\,k_1} \otimes \op{A}_{l_2\,k_2}) = \left\langle(\op{\mathcal J}_+^{2j_1-l_1+1} \op{\mathcal J}_-^{2j_1} \op{\mathcal J}_+^{k_1-1})\otimes(\op{\mathcal J}_+^{2j_2-l_2+1} \op{\mathcal J}_-^{2j_2} \op{\mathcal J}_+^{k_2-1})\right\rangle,
\label{density_bip}
\end{equation}
where $\tilde{k} = k_1k_2$ and $\tilde{l} = l_1l_2$, with $\tilde{k},\,\tilde{l} = 1,\ldots,d_1d_2$.

In the case of a separable density matrix expressed as a direct product $\op{\rho}=\op{\rho}_1 \otimes \op{\rho}_2$, the expression in Eq.~(\ref{density_bip}) will also correspond to the direct product of the expectation values
\begin{equation}
\rho_{\tilde{k}\tilde{l}}^{(sep)} = \left\langle\op{\mathcal J}_+^{2j_1-l_1+1} \op{\mathcal J}_-^{2j_1} \op{\mathcal J}_+^{k_1-1}\right\rangle\, \left\langle\op{\mathcal J}_+^{2j_2-l_2+1} \op{\mathcal J}_-^{2j_2} \op{\mathcal J}_+^{k_2-1}\right\rangle = \Tr{\left( \op{\rho}_1 \op{A}_{l_1k_1}\right)}  \Tr{\left( \op{\rho}_2 \op{A}_{l_2k_2}\right)}.
\end{equation}
In order to study the entanglement properties, we construct the reduced density matrices for each subsystem and the partial-transform density matrix elements. The reduced density matrices take the form
\begin{eqnarray}
\rho^{(1)}_{k_1l_1} = \sum_{k_2=1}^{d_2} \rho_{k_1k_2,l_1k_2} = \sum_{k_2=1}^{d_2} \left\langle(\op{\mathcal J}_+^{2j_1-l_1+1} \op{\mathcal J}_-^{2j_1} \op{\mathcal J}_+^{k_1-1})\otimes(\op{\mathcal J}_+^{2j_2-k_2+1} \op{\mathcal J}_-^{2j_2} \op{\mathcal J}_+^{k_2-1})\right\rangle, \nonumber \\
\rho^{(2)}_{k_2,l_2} =\sum_{k_1=1}^{d_1} \rho_{k_1k_2,k_1l_2} = \sum_{k_1=1}^{d_1} \left\langle(\op{\mathcal J}_+^{2j_1-k_1+1} \op{\mathcal J}_-^{2j_1} \op{\mathcal J}_+^{k_1-1})\otimes(\op{\mathcal J}_+^{2j_2-l_2+1} \op{\mathcal J}_-^{2j_2} \op{\mathcal J}_+^{k_2-1})\right\rangle,
\end{eqnarray}
and the partial transposed density matrix elements are
\begin{equation}
\rho_{\tilde{k}\tilde{l}}{}^{T_1} = \Tr(\op{\rho}\,\op{A}_{k_1\,l_1} \otimes \op{A}_{l_2\,k_2}), \quad \rho_{\tilde{k}\tilde{l}}{}^{T_2} = \Tr(\op{\rho}\,\op{A}_{l_1\,k_1} \otimes \op{A}_{k_2\,l_2})
\end{equation}
These expressions allow us to study both subsystems and the entanglement between them. An interesting example is the evaluation of the Clauser-Horne-Shimony-Holt (CHSH) inequality, which we do in the next Subsection for a two-qubit system and for a qubit-qutrit system.

\subsection{CHSH inequality for two--qubit entanglement.}

In the case of a two--qubit system one can express the total density matrix as the tensor product $\op{\rho} = \op{\rho}^{(1)}\otimes \op{\rho}^{(2)}$ in the following way
\begin{equation}
\op{\rho}=\left(
\begin{array}{cccc}
\langle \op{\sigma}_+ \op{\sigma}_- \otimes \op{\sigma}_+ \op{\sigma}_- \rangle & \langle \op{\sigma}_+ \op{\sigma}_- \otimes \op{\sigma}_- \rangle & \langle \op{\sigma}_- \otimes \op{\sigma}_+ \op{\sigma}_- \rangle & \langle \op{\sigma}_- \otimes \op{\sigma}_- \rangle\\
\langle \op{\sigma}_+ \op{\sigma}_- \otimes \op{\sigma}_+ \rangle & \langle \op{\sigma}_+ \op{\sigma}_- \otimes \op{\sigma}_- \op{\sigma}_+ \rangle & \langle \op{\sigma}_- \otimes \op{\sigma}_+ \rangle & \langle \op{\sigma}_- \otimes \op{\sigma}_- \op{\sigma}_+ \rangle \\
\langle \op{\sigma}_+ \otimes \op{\sigma}_+ \op{\sigma}_- \rangle & \langle \op{\sigma}_+ \otimes \op{\sigma}_- \rangle & \langle \op{\sigma}_-\op{\sigma}_+ \otimes \op{\sigma}_+ \op{\sigma}_- \rangle & \langle \op{\sigma}_- \op{\sigma}_+ \otimes \op{\sigma}_- \rangle \\
\langle \op{\sigma}_+ \otimes \op{\sigma}_+ \rangle & \langle \op{\sigma}_+ \otimes \op{\sigma}_- \op{\sigma}_+ \rangle & \langle \op{\sigma}_- \op{\sigma}_+ \otimes \op{\sigma}_+ \rangle & \langle \op{\sigma}_- \op{\sigma}_+ \otimes \op{\sigma}_- \op{\sigma}_+ \rangle
\end{array}\right),
\label{eqrho}
\end{equation}
where $\op{\sigma}_{\pm} = (\op{\sigma}_x \pm i\,\op{\sigma}_y)/2$ and in this case $\op{\mathcal{J}}_{\star} = \op{\sigma}_{\star}$. This expression may be used to evaluate the entanglement between the two subsystems. To this effect, one can use any measurement such as the logarithmic negativity~\cite{plenio05} or the CHSH inequality~\cite{clauser78}. In this work we explicitly present some violations of this inequality.

Given two different measurements by observers Alice and Bob performed by dichotomic observables $\op{A}_{\ell}$ and $\op{B}_{\ell}$ ($\ell = 1,2)$, respectively, one can construct the Clauser-Horne-Shimony-Holt (CHSH) inequality as
\begin{equation}
F_B := \vert E(A_1, B_1)+E(A_1,B_2)-E(A_2, B_1)+E(A_2,B_2) \vert \leq 2 \,,
\end{equation}
with $E(A_{\ell_1}, B_{\ell_2})={\rm Tr}(\op{\rho} \, (\op{A}_{\ell_1} \otimes \op{B}_{\ell_2}))$, and $F_B$ stands for the ``Bell parameter''.

Choosing
\begin{equation}
\op{A}_1=\op{\sigma}_z, \quad \op{A}_2=\op{\sigma}_x, \quad \op{B}_1=\frac{\op{\sigma}_x + \op{\sigma}_z}{\sqrt{2}}, \quad \op{B}_2=\frac{\op{\sigma}_x - \op{\sigma}_z}{\sqrt{2}}
\label{obs}
\end{equation}
and the two-qubit density matrix in terms of angular momentum operators given above, one arrives at
\begin{eqnarray}
&&\sqrt{2} \vert \langle \op{\sigma}_+ \op{\sigma}_- \otimes \op{\sigma}_- \rangle + \langle \op{\sigma}_+ \op{\sigma}_- \otimes \op{\sigma}_+ \rangle
+ \langle \op{\sigma}_- \otimes \op{\sigma}_- \op{\sigma}_+ \rangle+ \langle  \op{\sigma}_+ \otimes \op{\sigma}_- \op{\sigma}_+ \rangle \nonumber \\
&&-
 \langle \op{\sigma}_- \otimes \op{\sigma}_+ \op{\sigma}_- \rangle  - \langle  \op{\sigma}_+ \otimes \op{\sigma}_+ \op{\sigma}_- \rangle -
\langle \op{\sigma}_- \op{\sigma}_+ \otimes \op{\sigma}_- \rangle - \langle  \op{\sigma}_- \op{\sigma}_+ \otimes \op{\sigma}_+ \rangle
\vert \leq 2,
\label{chsh_2q}
\end{eqnarray}
which may be evaluated experimentally on a quantum system where the angular momentum degrees of freedom are dominant.

\begin{figure}
\centering
\includegraphics[width=0.45\linewidth]{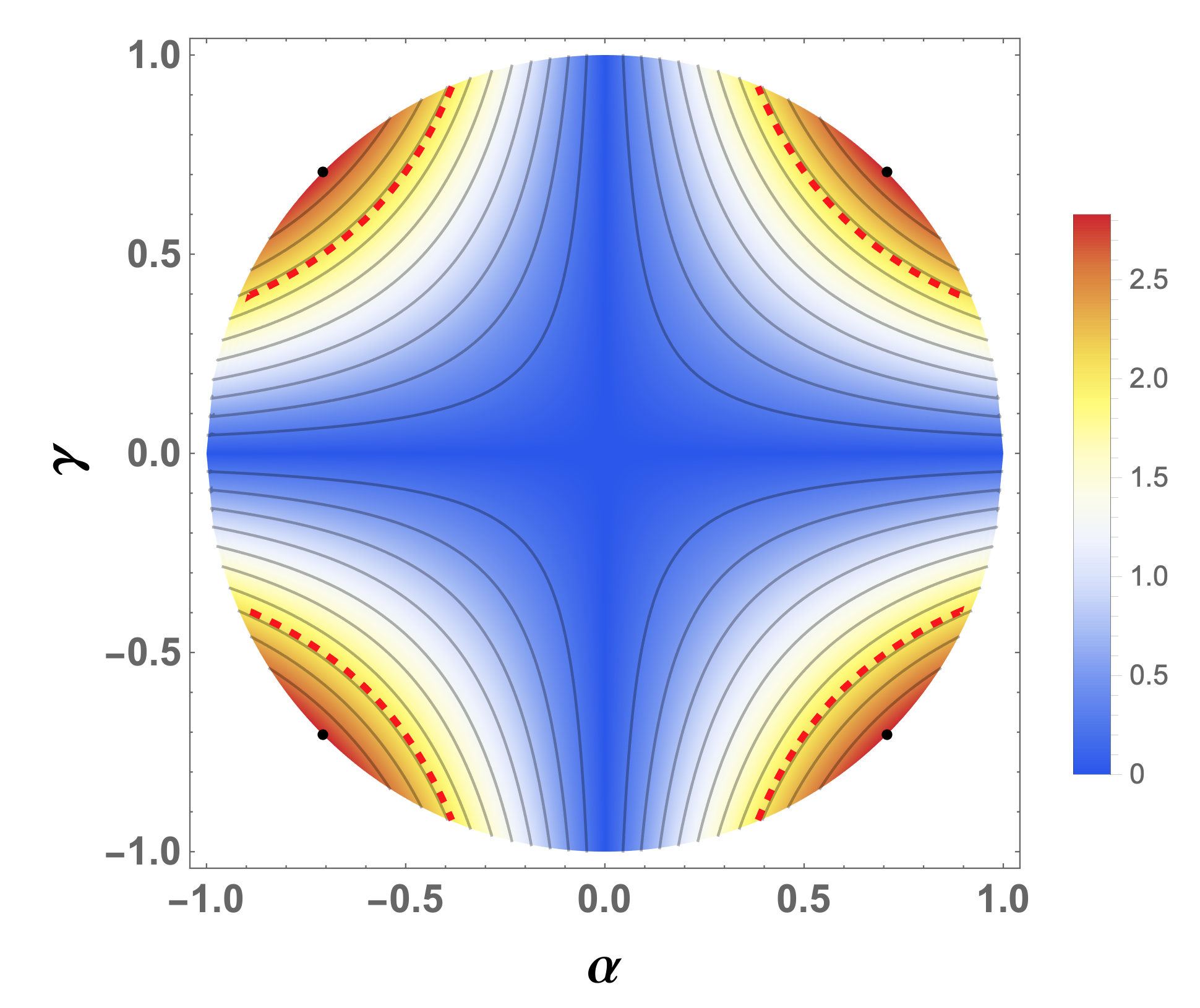}\quad
\includegraphics[width=0.45\linewidth]{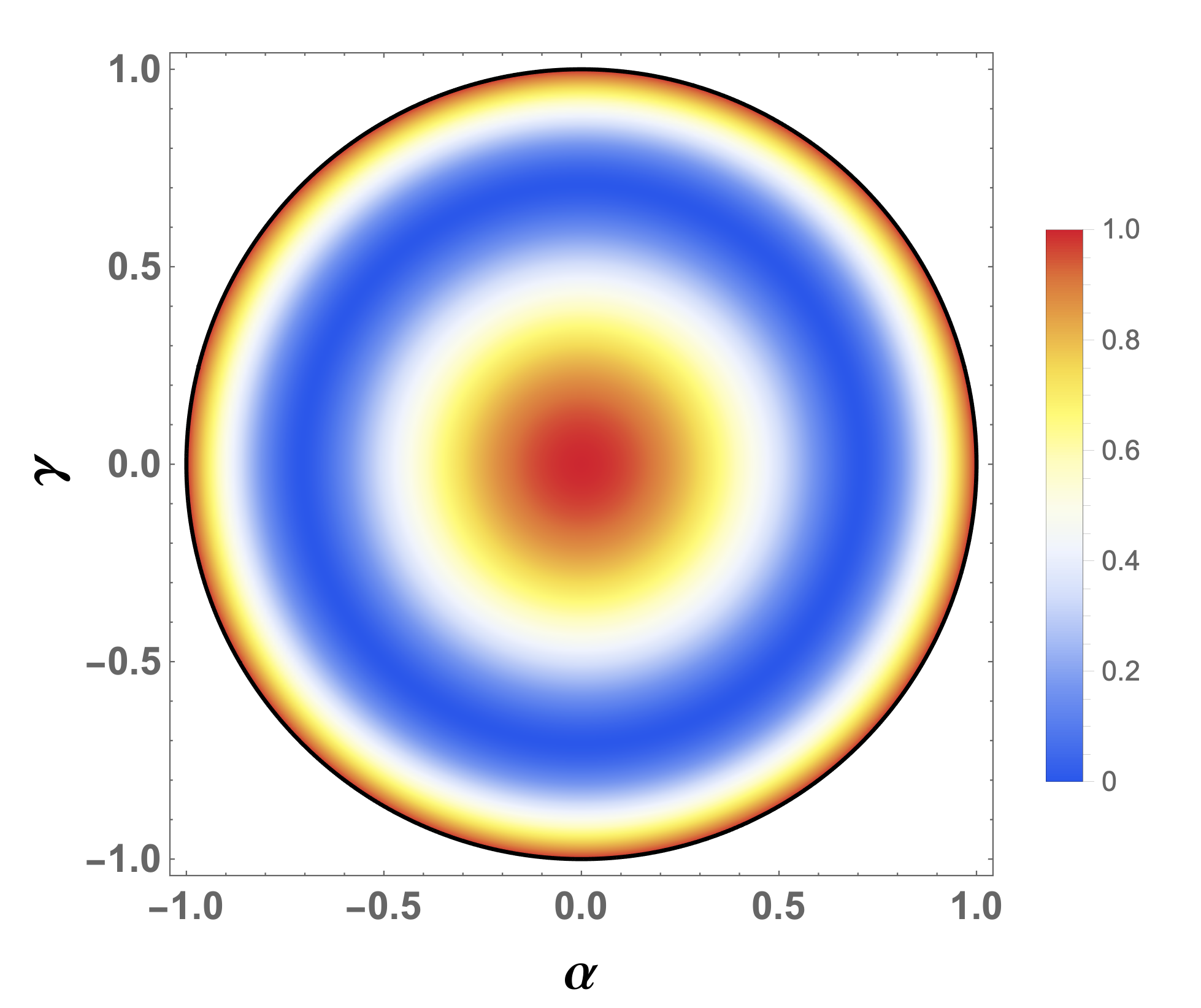}
\caption{Left: Contour plots for the inequality~(\ref{chsh_2q}) are shown in terms of the parameters $\alpha$ and $\gamma$ which define the state $\vert\Psi\rangle$ in eq.~(\ref{psi}), for the chosen observables~(\ref{obs}). The region of classical correlations, with $F_B\leq2$ is bounded by (red) dashed lines; beyond these boundaries the Bell parameter is $F_B>2$. The four (black) dots correspond to states for which the maximum value of $2\sqrt{2}$ is obtained. Right: Corresponding entanglement of formation for the same states (see text), and region (black, solid line) where it coincides with the Schlienz-Mahler $\beta$ parameter. The (red) line along the periphery corresponds to states with maximum violation of the CHSH inequality.}
\label{fig1}
\end{figure}

Notice that the inequality in~(\ref{chsh_2q}) may be studied for a general quantum state written in terms of the four two-qubit Bell states
\begin{equation}
\vert \phi_\pm \rangle= \frac{1}{\sqrt{2}}(\vert ++ \rangle \pm \vert -- \rangle), \quad \vert \psi_\pm \rangle = \frac{1}{\sqrt{2}}(\vert +- \rangle \pm \vert -+ \rangle).
\end{equation}
A general state of the form
\begin{equation}
\vert \Psi \rangle= \alpha \vert \phi_+ \rangle + \beta \vert \phi_- \rangle + \gamma \vert \psi_- \rangle + \delta \vert \psi_+ \rangle,
\label{psi}
\end{equation}
with $|\alpha|^2 + |\beta|^2 + |\gamma|^2 + |\delta|^2 = 1$, yields a CHSH inequality~(\ref{chsh_2q}) of the form
\begin{equation}
\left\vert 4 \sqrt{2} \alpha \gamma \right\vert \leq 2,
\end{equation}
which can indicate non-classical correlations when $\vert \alpha \gamma \vert > \frac{1}{2 \sqrt{2}}$, and leads to a maximal violation of the CHSH inequality when $\vert \alpha \gamma\vert = 1/2$. For the latter to be the case we have,
\begin{equation}
\vert \Psi_\pm \rangle=\frac{1}{\sqrt{2}}(\vert \phi_+\rangle \pm \vert \psi_- \rangle)\,,
\end{equation}
and $\pm \vert \Psi_\pm \rangle$ correspond to the maximal Bell number $F_B=2\sqrt{2}$. In Figure~\ref{fig1} (left) we show contour plots for the inequality~(\ref{chsh_2q}) in terms of $\alpha$ and $\gamma$, for all states of the form $\vert\Psi\rangle$ in eq.~(\ref{psi}) and for the chosen observables~(\ref{obs}). The region of classical correlations, with $F_B\leq2$ is bounded by (red) dashed lines; beyond these boundaries the Bell parameter is $F_B>2$. The four (black) dots correspond to the maximum value of $2\sqrt{2}$ obtained for the states $\vert \Psi_\pm \rangle$ (first and fourth quadrants) and $\pm \vert \Psi_\pm \rangle$ (second and third quadrants), respectively.
Figure~\ref{fig1} (right) shows the entanglement of formation for the same states $\vert \Psi \rangle$, defined, in terms of the concurrence $C$, by
\begin{equation}
E_F(\rho) = - x_+ \log_2(x_+) - x_- \log_2(x_-)
\end{equation}
where $x_{\pm} = (1 \pm \sqrt{1- C^2})/2$~\cite{wootters98}. The states on the periphery (red) correspond with the states of maximum violation of the CHSH inequality; here, the probability of finding the system in the states $\vert\phi_-\rangle$ and $\vert\psi_+\rangle$ vanishes. The central region, where the contribution of $\vert\phi_-\rangle$ and $\vert\psi_+\rangle$ is greater, shows also a maximum value for the entanglement of formation. Furhermore, the plot of the Schlienz-Mahler $\beta$ parameter~\cite{schlienz95} as a function of $\alpha$ and $\gamma$ has the same shape as that of the entanglement of formation; the solid black line at the external border shows the region where these two are equal.

The entanglement of formation of real linear combinations of Bell states has been studied together with its relevance in quantum correction codes and in entanglement purification protocols. In particular, the entanglement properties cannot change by making local rotations in the basis states~\cite{bennett96}.
Here, working in the Bell basis amounts to a rotation with respect to the computational basis; Figure~\ref{fig1} remains the same but now the axes would be $\alpha + \beta$ and $\gamma - \delta$ and the Bell factor takes the form $F_B = \vert 4\sqrt{2}(\alpha+\beta)(\gamma - \delta)\vert$.

Notice that the commuting element $\op{X}_1$~(cf. eq.(\ref{eqX})) can be used to generate the most general X-state from the density matrix of two qubits
\begin{equation}
\op{\rho}_{X} =\frac{1}{2}\left( \op{\rho} + \op{X}_1 \op{\rho} \op{X}_1\right) =
\left(
\begin{array}{cccc}
 \rho_{11}&0  & 0 & \rho_{14} \\
 0 & \rho_{22}  & \rho_{23} & 0 \\
 0  & \rho_{32}  & \rho_{33} & 0  \\
 \rho_{41}  & 0  & 0 &  \rho_{44} 
\end{array} 
\right) \, ,
\label{Xstate}
\end{equation}
with $\rho_{jk}$ the matrix elements of $\op{\rho}$.

\begin{figure}
\centering
\includegraphics[width=0.40\linewidth]{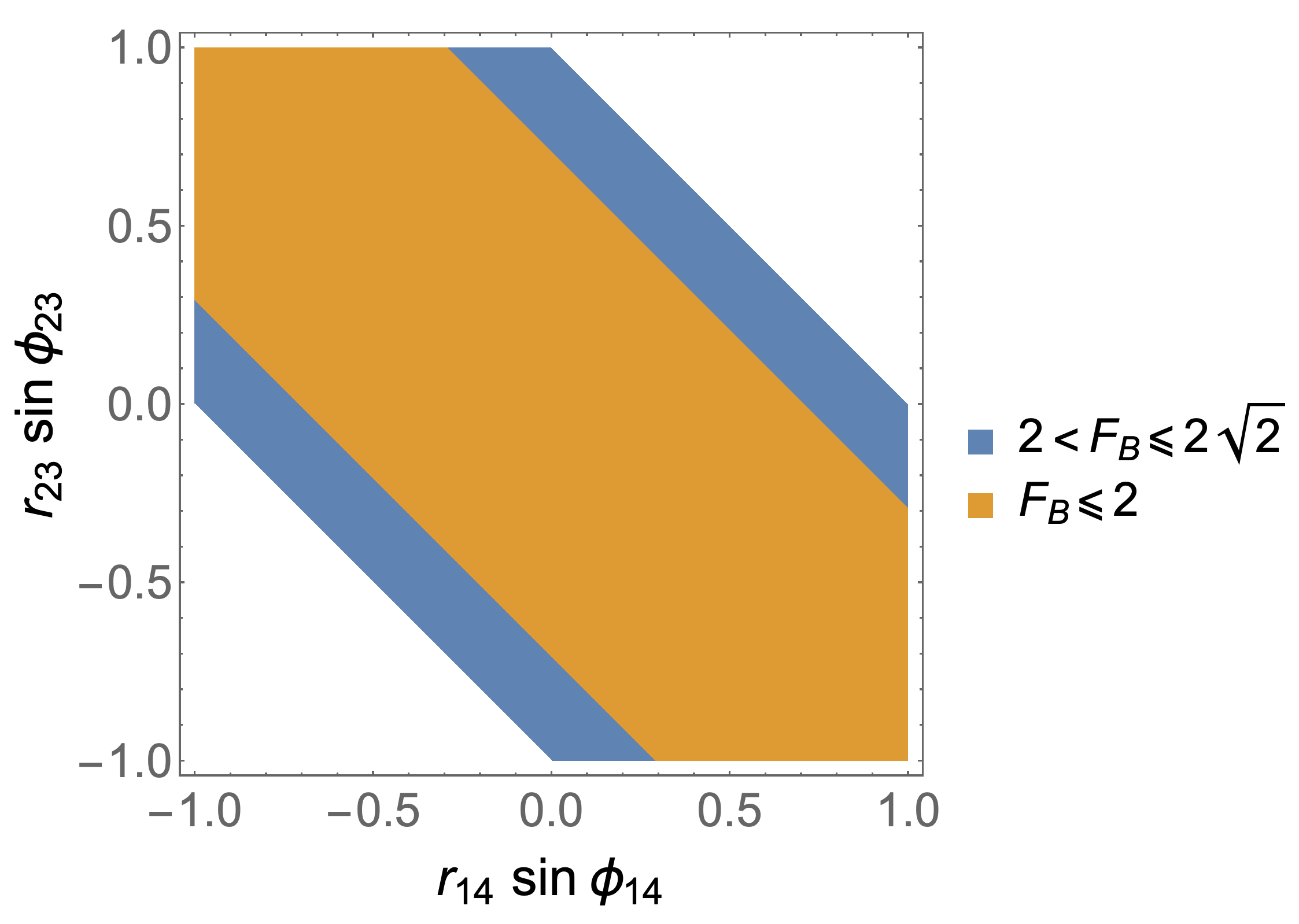}\quad
\includegraphics[width=0.45\linewidth]{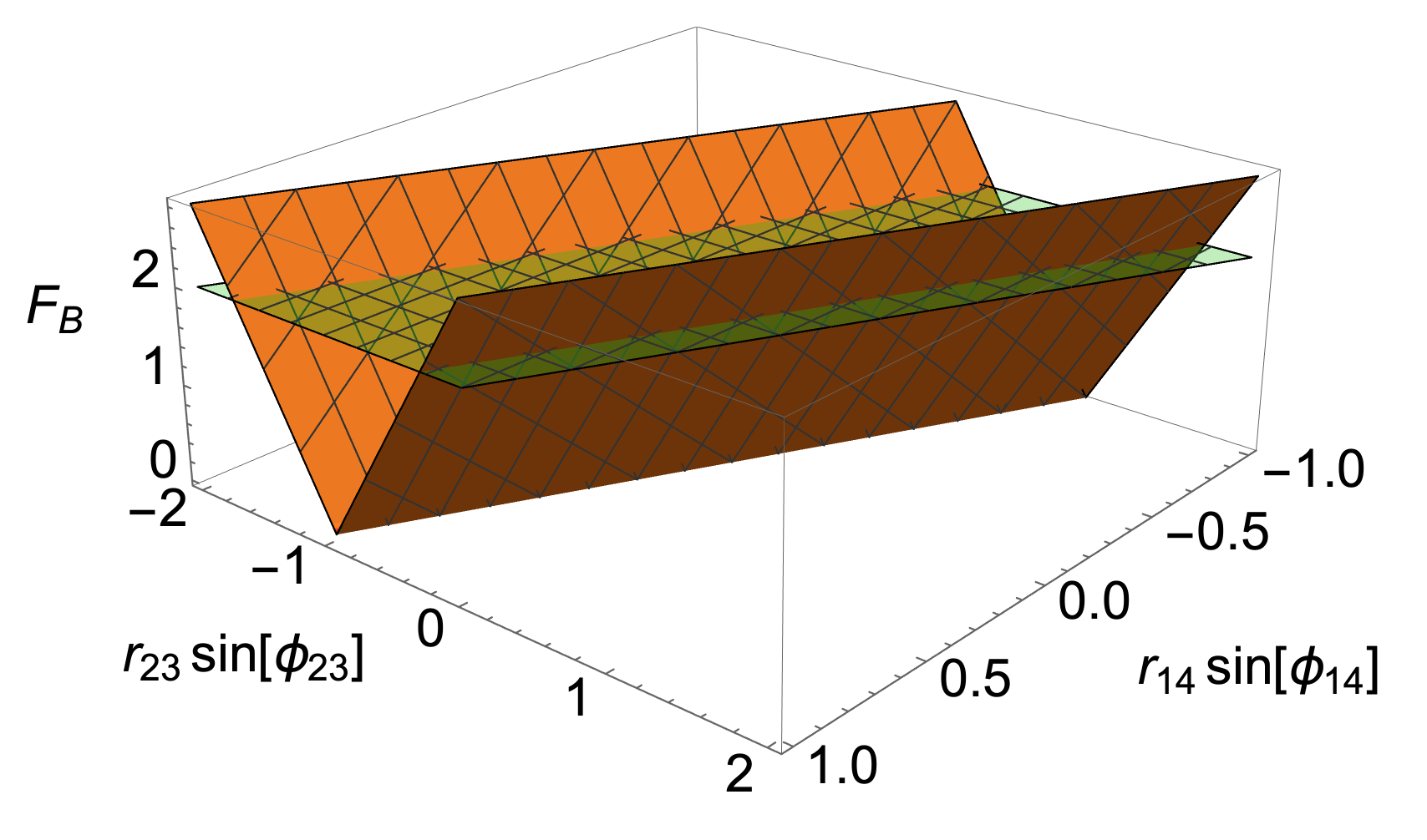}
\caption{Left: Contour plot of the Bell parameter as a function of $r_{14} \sin\phi_{14}$ and $r_{23} \sin\phi_{23}$ for the X-state~(\ref{Xstate}) with observables~(\ref{obs2}). The regions indicate where the CHSH inequality is satisfied or violated. Along the boundary the Cirel'son limit is attained. Right: $3$-dimensional plot of the same. The boundary value $F_B$=2 is indicated by a horizontal (green) plane.}
\label{fig2}
\end{figure}

For this X-state we select the operators 
\begin{equation}
\op{A}_1=\op{\sigma}_y, \quad \op{A}_2=\op{\sigma}_x, \quad \op{B}_1=\frac{\op{\sigma}_x + \op{\sigma}_z}{\sqrt{2}}, \quad \op{B}_2=\frac{\op{\sigma}_x - \op{\sigma}_z}{\sqrt{2}}\,,
\label{obs2}
\end{equation}
and the Bell factor takes the form
\begin{equation}
F_B=2\sqrt{2}\,\left| r_{14} \sin\phi_{14} + r_{23} \sin\phi_{23} \right|\,,
\end{equation}
where we define 
\begin{eqnarray*}
\rho_{14} &=& \langle \op\sigma_+ \otimes \op\sigma_+ \rangle =: r_{14} \, e^{i \phi_{14}}\,, \\
\rho_{23} &=& \langle \op\sigma_+ \otimes \op\sigma_- \rangle =: r_{23} \, e^{i \phi_{23}}\,.
\end{eqnarray*}
Figure~\ref{fig2} shows (left) a contour plot of the Bell parameter as a function of the imaginary part of the expectation values of $\rho_{14}$ and $\rho_{23}$, which appear in the expression for $F_B$, respectively   $r_{14} \sin\phi_{14}$ and $r_{23} \sin\phi_{23}$; and (right) the corresponding three dimensional plot. In both cases the regions where $F_B$ is smaller than or equal to $2$, and where it is greater than $2$ are clearly indicated. The Cirel'son limit  $F_B=2 \sqrt 2$ is also reached for a set of values of $(r_{14}, r_{23})$. 
%

\subsection{CHSH inequality and qubit--qutrit entanglement.}

In a qubit--qutrit interaction, by using the angular momentum representation of the density matrix the state is given by a $6\times 6$ matrix which can be expressed as 

{ \footnotesize
\begin{equation}
\resizebox{0.93\hsize}{!}{$
\op{\rho} =
\left(
\begin{array}{cccccc}
\langle \op{\sigma}_+ \op{\sigma}_- \otimes \opcal{ J}_+^2 \opcal{ J}_-^2  \rangle & \langle \op{\sigma}_+ \op{\sigma}_- \otimes \opcal{ J}_+ \opcal{ J}_-^2 \rangle & \langle \op{\sigma}_+ \op{\sigma}_- \otimes \opcal{ J}_-^2\rangle & \langle \op{\sigma}_- \otimes \opcal{ J}_+^2 \opcal{ J}_-^2  \rangle & \langle \op{\sigma}_- \otimes \opcal{ J}_+ \opcal{ J}_-^2 \rangle & \langle \op{\sigma}_- \otimes \opcal{ J}_-^2\rangle \\
\langle \op{\sigma}_+ \op{\sigma}_- \otimes \opcal{ J}_+^2 \opcal{ J}_-\rangle & \langle \op{\sigma}_+ \op{\sigma}_- \otimes \opcal{ J}_- \opcal{ J}_+^2 \opcal{ J}_- \rangle & \langle \op{\sigma}_+ \op{\sigma}_- \otimes \opcal{ J}_-^2 \opcal{ J}_+\rangle & \langle \op{\sigma}_- \otimes \opcal{ J}_+^2 \opcal{ J}_-\rangle & \langle \op{\sigma}_- \otimes \opcal{ J}_- \opcal{ J}_+^2 \opcal{ J}_- \rangle & \langle \op{\sigma}_- \otimes \opcal{ J}_-^2 \opcal{ J}_+\rangle\\
\langle \op{\sigma}_+ \op{\sigma}_- \otimes \opcal{ J}_+^2\rangle & \langle \op{\sigma}_+ \op{\sigma}_- \otimes \opcal{ J}_- \opcal{ J}_+^2\rangle & \langle \op{\sigma}_+ \op{\sigma}_- \otimes \opcal{ J}_-^2 \opcal{ J}_+^2 \rangle & \langle \op{\sigma}_- \otimes \opcal{ J}_+^2\rangle & \langle \op{\sigma}_- \otimes \opcal{ J}_- \opcal{ J}_+^2\rangle & \langle \op{\sigma}_- \otimes \opcal{ J}_-^2 \opcal{ J}_+^2 \rangle \\
\langle \op{\sigma}_+ \otimes \opcal{ J}_+^2 \opcal{ J}_-^2  \rangle & \langle \op{\sigma}_+ \otimes \opcal{ J}_+ \opcal{ J}_-^2 \rangle & \langle \op{\sigma}_+ \otimes \opcal{ J}_-^2\rangle & \langle \op{\sigma}_- \op{\sigma}_+ \otimes \opcal{ J}_+^2 \opcal{ J}_-^2  \rangle & \langle \op{\sigma}_- \op{\sigma}_+ \otimes \opcal{ J}_+ \opcal{ J}_-^2 \rangle & \langle \op{\sigma}_- \op{\sigma}_+ \otimes \opcal{ J}_-^2\rangle \\
\langle \op{\sigma}_+ \otimes \opcal{ J}_+^2 \opcal{ J}_-\rangle & \langle \op{\sigma}_+ \otimes \opcal{ J}_- \opcal{ J}_+^2 \opcal{ J}_- \rangle & \langle \op{\sigma}_+ \otimes \opcal{ J}_-^2 \opcal{ J}_+\rangle & \langle \op{\sigma}_- \op{\sigma}_+ \otimes \opcal{ J}_+^2 \opcal{ J}_-\rangle & \langle \op{\sigma}_- \op{\sigma}_+ \otimes \opcal{ J}_- \opcal{ J}_+^2 \opcal{ J}_- \rangle & \langle \op{\sigma}_- \op{\sigma}_+ \otimes \opcal{ J}_-^2 \opcal{ J}_+\rangle \\
\langle \op{\sigma}_+ \otimes \opcal{ J}_+^2\rangle & \langle \op{\sigma}_+ \otimes \opcal{ J}_- \opcal{ J}_+^2\rangle & \langle \op{\sigma}_+ \otimes \opcal{ J}_-^2 \opcal{ J}_+^2 \rangle & \langle \op{\sigma}_- \op{\sigma}_+ \otimes \opcal{ J}_+^2\rangle & \langle \op{\sigma}_- \op{\sigma}_+ \otimes \opcal{ J}_- \opcal{ J}_+^2\rangle & \langle \op{\sigma}_- \op{\sigma}_+ \otimes \opcal{ J}_-^2 \opcal{ J}_+^2 \rangle
\end{array}
\right),
$}
\label{eq.38}
\end{equation}
}
\noindent where $\opcal{ J}_\pm = \op{J}_\pm /\sqrt{2}$.

For a quantum system of two particles with angular momenta $j_1$ and $j_2$ as the main degrees of freedom, the Hamiltonian is given by
\begin{equation}
H =  2 \, \omega_0 \op{J}_1 \cdot \op{J}_2 - \omega_1 \op{J}_z \,,
\end{equation}
with $\op{J}_z = \op{J}_{1 z} + \op{J}_{2 z}$ and $\omega_1,\, \omega_0$ denote the frequency parameter strengths, with $\omega_1 << \omega_0$. The energy spectrum can be obtained in the representation in which the total angular momentum $\op{J}^2$ and its component $\op{J}_z$ as well as $\op{J}^2_1$ and $\op{J}^2_2$ are diagonal with eigenvalues $j$ and $m$. This coupled representation $| j_1, j_2; j m\rangle$ is connected with the uncoupled representation $|j_1, m_1\rangle | j_2, m_2\rangle$ by a unitary transformation,
where the elements of the transformation $\langle j_1, m_1 , j_2, m-m_1 | j ,m\rangle$ are the Clebsch-Gordan coefficients~\cite{rose11, messiah81}.

Thus, the energy of the system is given by,
\begin{equation}
E(j_1,j_2,j,m) =  \omega_0 \left\{ j (j +1) - j_1 (j_1 +1)- j_2(j_2+1) \right\}  - \omega_1 m \, ,
\label{eqqE}
\end{equation}
where $m=-j,-j+1,\cdots, j$, and the state function is
\begin{equation}
|\varphi_{j_1 j_2 j m}(t)\,\rangle = \sum_{m_1} \langle j_1, m_1 , j_2, m-m_1 | j , m\rangle\, e^{-i \op{H}t}\, | j_1, m_1 \rangle | j_2, m-m_1 \rangle \, .
\end{equation}

When, as here, the case of two particles with angular momenta $j_1=1/2$ and $j_2=1$ is considered, the coupled total angular momentum can take the values $j=1/2,\, 3/2$; in this case the eigenstates can be written in terms of the computational (uncoupled) basis as follows,
\begin{eqnarray}
  |\varphi_{1/2, 1; 3/2, 3/2}(t)\,\rangle &=& e^{-i \omega_0 t}\,e^{i\frac{3}{2}\omega_1t} \, \left(1,0,0,0,0,0\right)^T , \nonumber \\[2mm]
  |\varphi_{1/2, 1; 3/2, 1/2}(t)\,\rangle&=& e^{-i \omega_0 t}\,e^{i\frac{1}{2}\omega_1t} \, \left(0,\sqrt{2/3} ,0,\sqrt{1/3},0,0\right)^T \, ,\nonumber  \\[2mm]
  |\varphi_{1/2, 1; 3/2, -1/2}(t)\,\rangle &=& e^{-i \omega_0 t}\,e^{-i\frac{1}{2}\omega_1t} \, \left(0,0,\sqrt{1/3} ,0,\sqrt{2/3},0\right)^T \,, \nonumber  \\[2mm]
  |\varphi_{1/2, 1; 3/2, -3/2}(t)\,\rangle &=& e^{-i \omega_0 t}\,e^{-i\frac{3}{2}\omega_1t} \,  \left(0,0,0,0,0,1\right)^T\, , \nonumber \\[2mm]
  |\varphi_{1/2, 1; 1/2, 1/2}(t)\,\rangle &=& e^{2i \omega_0 t}\,e^{i\frac{1}{2}\omega_1t} \, \left(0,\sqrt{1/3} ,0,-\sqrt{2/3},0,0\right)^T  \nonumber \\
  |\varphi_{1/2, 1; 1/2, -1/2}(t)\,\rangle &=& e^{2i \omega_0 t}\,e^{-i\frac{1}{2}\omega_1t} \, \left(0,0,\sqrt{2/3} ,0,-\sqrt{1/3},0\right)^T \, .
\end{eqnarray}

In order to study the behavior of the Bell inequality, linear combinations of the eigenstates are used and we may omit global phases; using a dimensionless time parameter $\tau = \omega_1t$, we have
\begin{eqnarray}
| \Psi_1(\tau)\rangle &=& \left(e^{i\frac{3}{2}\tau}\, \cos\theta_1,0,0,0,0, e^{-i\frac{3}{2}\tau}\, \sin\theta_1\right)^T , \nonumber \\[2mm]
| \Psi_2(\tau)\rangle &=& \, \left(0,e^{i\frac{1}{2}\tau}\, \sqrt{1/3} \cos\theta_2, e^{-i\frac{1}{2}\tau}\, \sqrt{2/3}  \sin\theta_2,-e^{i\frac{1}{2}\tau}\, \sqrt{2/3} \cos\theta_2,-e^{-i\frac{1}{2}\tau}\, \sqrt{1/3} \sin\theta_2,0\right)^T \, , \nonumber \\[2mm]
 | \Psi_3(\tau)\rangle &=& \left(0,e^{i\frac{1}{2}\tau}\, \sqrt{2/3} \cos\theta_3 ,e^{-i\frac{1}{2}\tau}\, \sqrt{1/3} \sin\theta_3,e^{i\frac{1}{2}\tau}\, \sqrt{1/3} \cos\theta_3,e^{-i\frac{1}{2}\tau}\, \sqrt{2/3} \sin\theta_3,0\right)^T  \, .
\end{eqnarray}
Next we consider a convex sum of the corresponding density matrices of the pure states $|\Psi_k\rangle$ with $k=1,2,3$, giving a density operator state
\begin{equation}
\rho(\tau) = p_1 | \Psi_1(\tau)\rangle\langle \Psi_1(\tau)| + p_2 | \Psi_2(\tau)\rangle\langle \Psi_2(\tau) | + p_3  | \Psi_3(\tau)\rangle\langle \Psi_3(\tau) | \, ,
\label{eq.rhoT}
\end{equation}
where $p_k$ denotes the probability to measure the corresponding pure state $|\Psi_k(\tau)\rangle$ with $k=1,2,3$, under the condition $p_1+p_2+p_3=1$.

\begin{figure}
\includegraphics[scale=0.3]{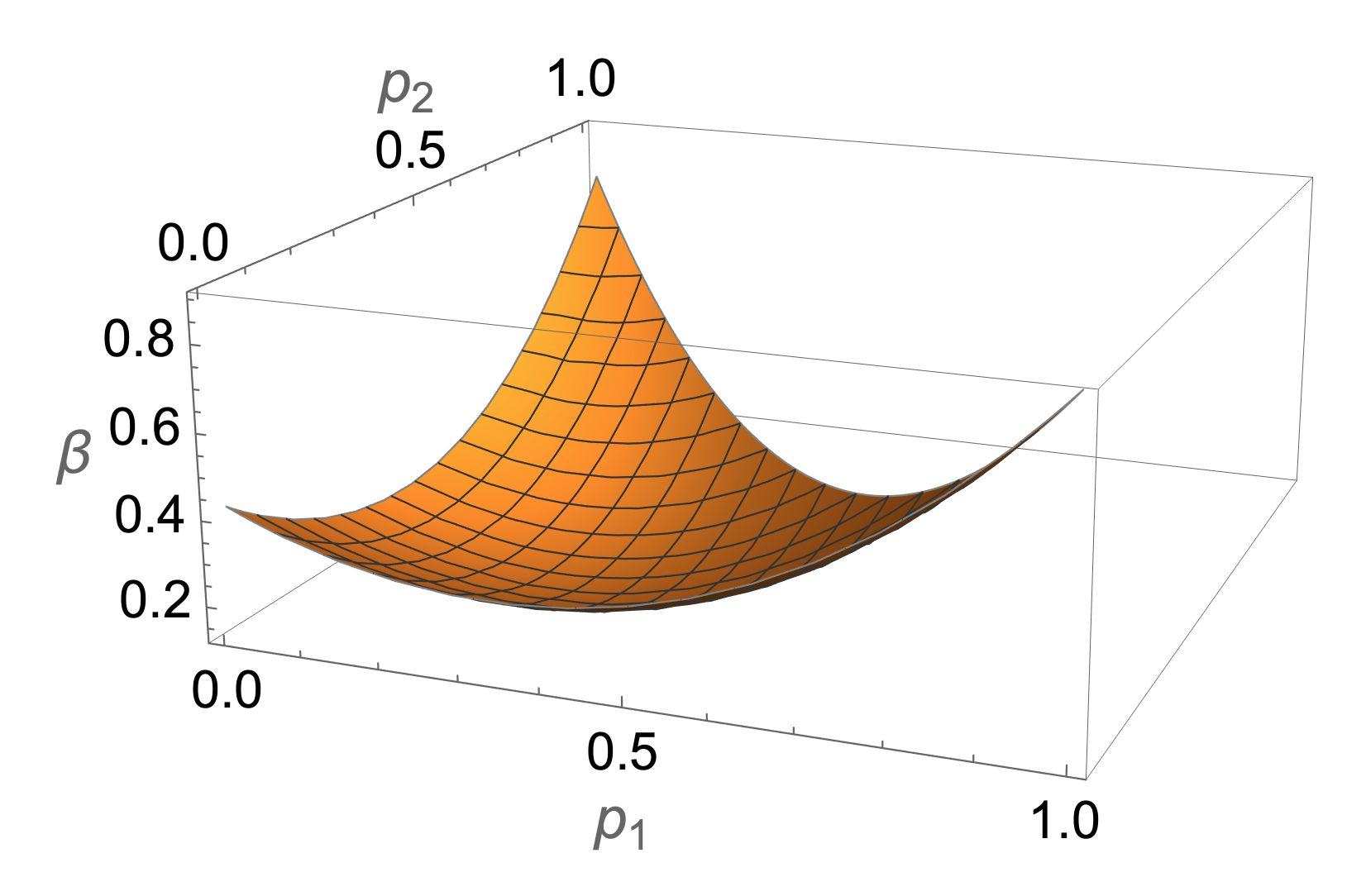} \quad 
\includegraphics[scale=0.4]{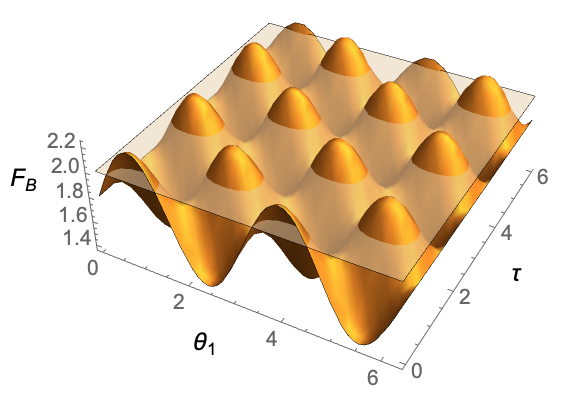} \\
\includegraphics[scale=0.42]{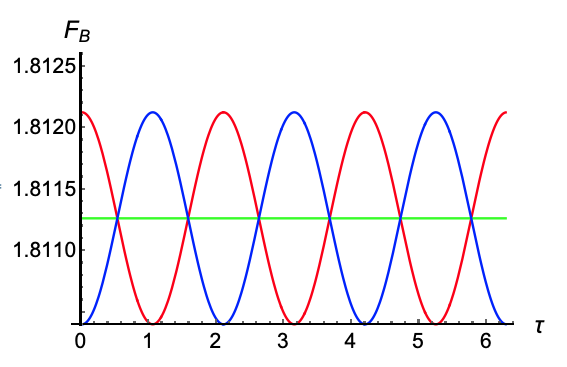} \qquad
\includegraphics[scale=0.275]{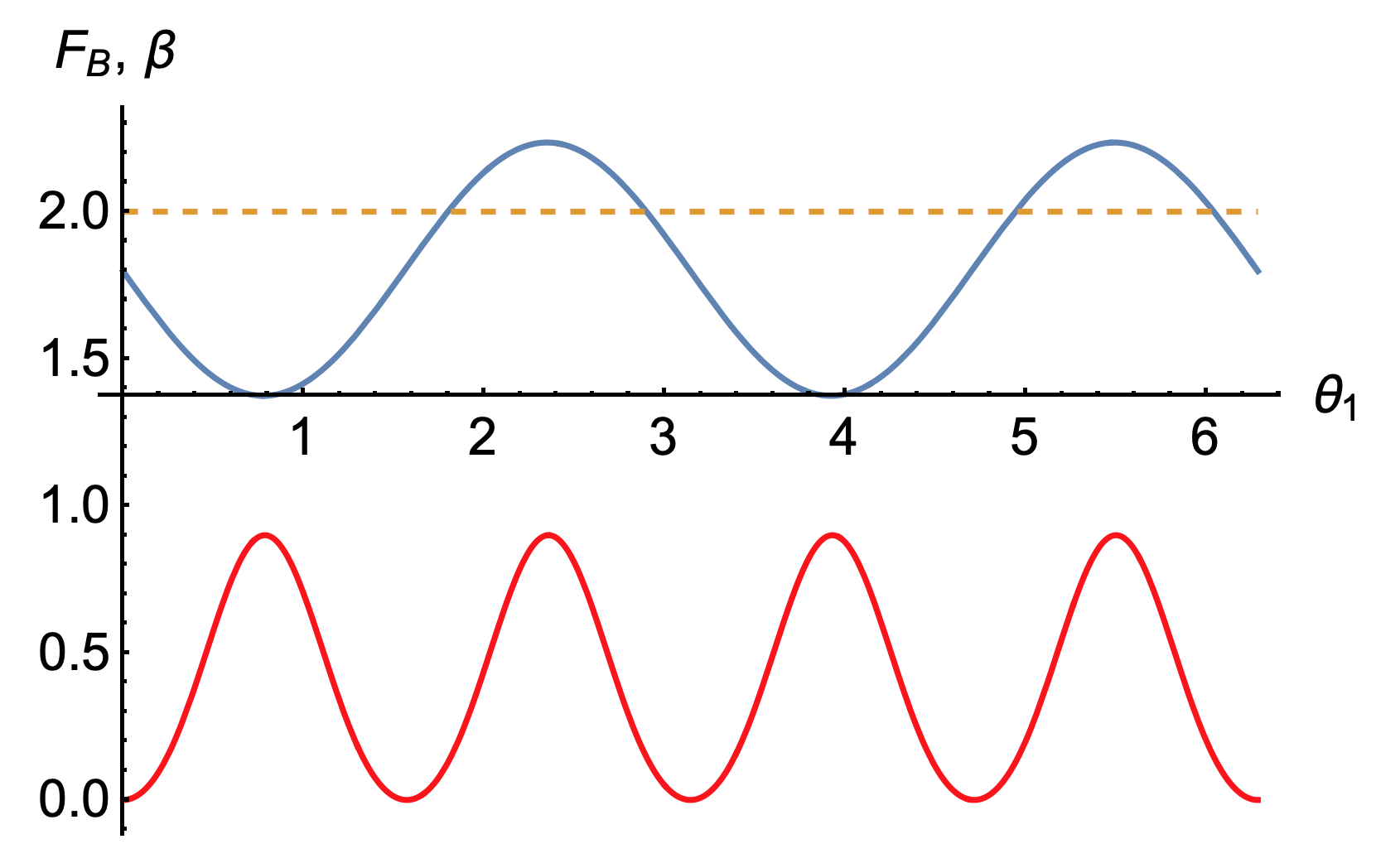}
\caption{Top left: $\beta$ entanglement measure as function of $p_1$ and $p_2$ for the $\rho$-state given in Eq.~(\ref{eq.rhoT}). Top right: the corresponding violation of Bell's inequality as a function of $\tau$ and $\theta_1$ when $p_1=1$, where the $\beta$ parameter is maximum. The plane $F_B=2$ is shown as reference. Bottom left: In the special case $\theta_1 = 3\pi/2$ the Bell parameter $F_B$ is constant. A slight variation of $\pm 10^{-3}$ in $\theta_1$ produces large variations in $F_B$. Bottom right: Both $F_B$ and the $\beta$-parameter are shown as functions of $\theta_1$, for $\tau = \pi$ and the parameters given in~(\ref{set1}). $\beta$ varies in the interval $[0,\,1]$ and its maxima coincide with the maxima and minima of $F_B$, which overflows above the Bell boundary of $2$.}
\label{fig3}
\end{figure}

To determine the entanglement properties of $\rho$ in the computational basis, we evaluate the Schlienz-Mahler $\beta$ parameter of entanglement for a qubit-qutrit system~\cite{schlienz95}. This takes the values $0\leq \beta\leq 1$. It is plotted in Figure~\ref{fig3} (top left) as a function of the probabilities $p_1$ and $p_2$ mentioned above, for $\tau =0$ in the density operator state~(\ref{eq.rhoT}). Notice that the maximum entanglement takes place at the borders  $p_1 \approx 1$ and $p_2 \approx 1$. As can be seen, the case $p_1= p_2 \approx 0$ also represents an entangled state, though with a smaller $\beta$ value than for the previous cases. From here onwards we consider the cases $p_1=1$ or $p_2=1$.

We shall consider then the case $p_1=1$ for the qubit-qutrit interaction, which yields an $X$-state of the form [from eq.(\ref{eq.38})]
\begin{equation}
\op{\rho} = \left(
\begin{array}{cccccc}
\cos^2\theta_1 & 0 & 0 & 0 & 0 & e^{i3\tau}\cos\theta_1\, \sin\theta_1 \\
0 & 0 & 0 & 0 & 0 & 0 \\
0 & 0 & 0 & 0 & 0 & 0 \\
0 & 0 & 0 & 0 & 0 & 0 \\
0 & 0 & 0 & 0 & 0 & 0 \\
 e^{-i3\tau}\cos\theta_1\, \sin\theta_1 & 0 & 0 & 0 & 0 & \sin^2\theta_1
\end{array}
\right) .
\label{xx}
\end{equation}

The observables for Alice and Bob are chosen as
\begin{eqnarray}
\op{A}_1 &=& \op{u}_1 \op{\sigma}_{1 z} \op{u}^\dagger_1  \, ,  \quad \op{A}_2 = \op{u}_2 \op{\sigma}_{1 z} \op{u}^\dagger_2 \, , \quad 
\op{B}_1 = \op{I}_3 - \op{J}_{2 z} (\op{J}_{2 z} -1)  \, , \quad \op{B}_2 = \op{U} \op{B}_1 \op{U}^\dagger \, ,
\end{eqnarray}
where the unitary transformations $\op{u}_1, \op{u}_2$ and $\op{U}$ are given by 
\begin{equation}
\op{u}_1 = e^{i (\alpha_1 \, \op{\sigma}_{1 z} + \alpha_2 \, \op{\sigma}_{1 y}) } \, , \quad  \op{u}_2 = e^{i (\alpha_3 \,\op{\sigma}_{1 z} + \alpha_4 \,\op{\sigma}_{1 y} + \alpha_5 \,\op{\sigma}_y) } \, , \quad \op{U}= e^{i \beta_1\, \op{J}_{2 z}} \, e^{i \beta_2 \, \op{J}_{2 x}} \, .
\end{equation}
We have $7$ free parameters available to fit the value of $F_B$. For the following set of parameters
\begin{equation}
\alpha_1=\frac{1}{2} \, , \alpha_2 = \frac{9}{2}\, , \alpha_3 = 5 \, , \alpha_4 = 1\, , \alpha_5= \frac{17}{4}\, ,  \beta_1 = \frac{4}{3}  \, , \beta_2= \pi \, .
\label{set1}
\end{equation}
one gets a violation of Bell's inequality with $F_B = 2.23503$ (see plots on the right in Fig.~\ref{fig3}).

In Fig~\ref{fig3}  the regions where the violation of Bell's inequality takes place are also shown as a function of $\tau$ and $\theta_1$, which are indicated above the plane $F_B=2$. The bottom plots show (left) the special case $\theta_1 = 3\pi/2$ for which Bell parameter $F_B$ is constant. Slightly varying $\theta_1 =  3\pi/2\pm 10^{-3}$ in $\theta_1$ produces variations in $F_B$. When $F_B$ and the $\beta$-parameter are plotted as functions of $\theta_1$, for $\tau = \pi$ and the parameters given in~(\ref{set1}) we see that $\beta$ reaches its maxima at the maxima and minima of $F_B$, which overflows above the Bell boundary of $2$.

%
%
%
\begin{figure}
\includegraphics[scale=0.4]{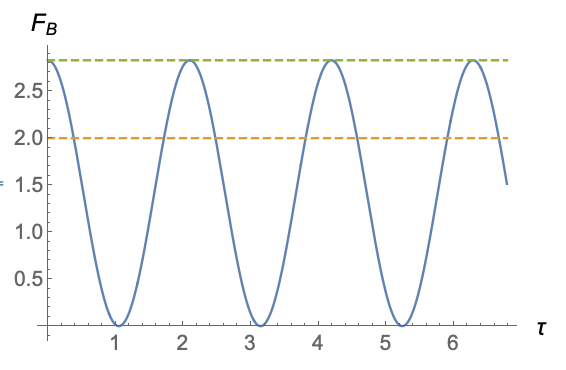} \qquad
\includegraphics[scale=0.3]{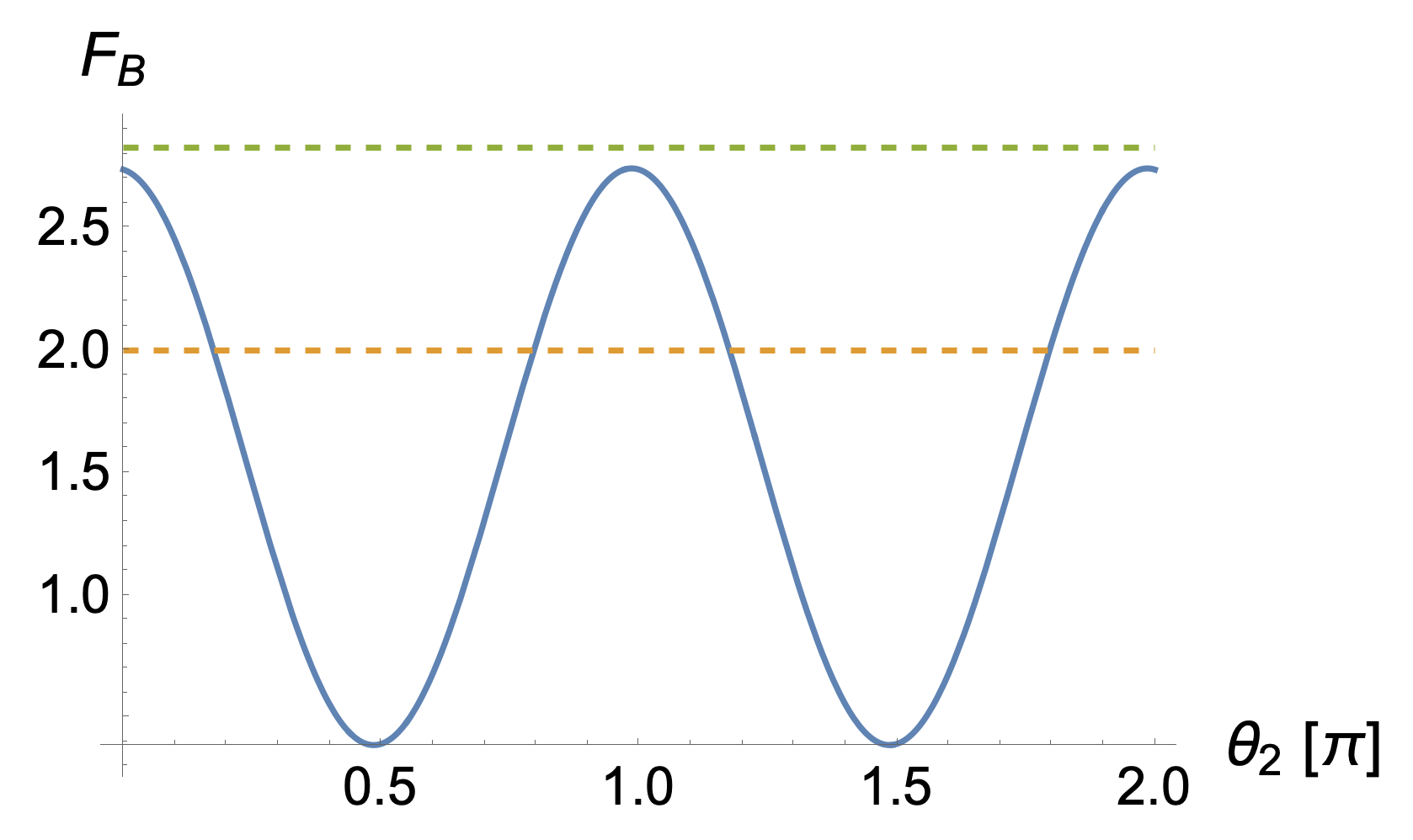}
\caption{Left: Bell parameter as a function of $\tau$ when $\theta_1 = 3\pi/4$, for the $X$-state of Eq.~(\ref{xx}) with $p_1=1$. Right: For the case  $p_2=1$ the Bell parameter is independent of $\tau$; it is here plotted as a function of $\theta_2$. Both cases show strong violations of Bell's inequality, close to the maximum limit established by Cirel'son's limit. In both cases the dashed lines mark the region where the inequality is violated.}
\label{FBX}
\end{figure}

It is possible to obtain a larger violation of the CHSH inequality by using the same density matrix~(\ref{xx}), if we take observables for Alice and Bob given by,
\begin{eqnarray}
\op{A}_1 = \left(
\begin{array}{cc}
 \cos \alpha &
 -\sin \alpha \\
-\sin \alpha &  -\cos \alpha \\
\end{array}
\right) , \qquad
\op{A}_2 = \left(
\begin{array}{cc}
 \cos c &
   -\sin c \\
  -\sin c &
   -\cos c \\
\end{array}
\right) , \nonumber \\
\op{B}_1 ={\rm Diag}(1,-1,-1), \quad \op{B}_2 = \op{U} \,{\rm Diag}(-1,1,1)\, \op{U}^\dagger, \quad \op{U} = \prod_{k=1}^6 \exp\left(i  b_{k} \op{\Lambda}_k \right),
\end{eqnarray}
where $U$ is a unitary transformation with $\op{\Lambda}_k$ ($k=1,2,\cdots 6$) denoting the non-diagonal generators of $\mathfrak{su}(3)$. In this case we have $8$ free parameters to find maximum values for the Bell factor. Using the Bell correlations and maximizing the Bell factor $F_B$ we obtain
\begin{equation}
\alpha = \pi/4,\ c = 3\pi/4,\ b_1=-\frac{\pi}{30},\ b_2=\frac{\pi}{2},\ b_3=0,\ 
b_4=\frac{3\pi}{4},\ b_5=\sqrt{\frac{3}{5}}\,\pi,\ b_6=\frac{\pi}{40},
\end{equation}
yielding
\begin{equation}
2\sqrt{2} - F_B \approx 5.58 \times 10^{-4}, \quad ({\rm for }\ \theta = 3\pi/4 ),
\end{equation}
i.e., $0.9998$ times the maximum value of $2\sqrt{2}$ corresponding to Cirel'son limit~\cite{cirelson80}.
By slightly varying the given values of the parameters, we may obtain the maximum violation.

As another example, for the state~(\ref{eq.rhoT}) with $p_1=0,\ p_2=1,\ p_3=0$, and parameters
\begin{equation}
\alpha = \pi/4,\ c = 3\pi/4,\ b_1=-\frac{\pi}{125},\ b_2=-b_1,\ b_3=\frac{\pi}{2},\ 
b_4=\frac{7\pi}{8},\ b_5=\frac{\pi}{5},\ b_6=-\frac{3\pi}{10},
\end{equation}
yields $F_B = 2.739$, i.e., $0.968$ times the Cirel'son limit. Figure~\ref{FBX} shows the behaviour of $F_B$ as a function of $\tau$ when $\theta_1 = 3\pi/4$, for the $X$-state of Eq.~(\ref{xx}) with $p_1=1$ (left). For the case  $p_2=1$ the Bell parameter is independent of $\tau$; it is here displayed (right) as a function of $\theta_2$. Both cases show strong violations of Bell's inequality, close to the maximum limit established by Cirel'son's limit. The upper bound and the maxima where $F_B$ is attained are also shown.

\section{Summary and concluding remarks}

The parametrization of the density matrix for any finite system was obtained by representing the generalized Gell-Mann matrices in terms of products of the angular momentum operators $\op J_\pm$.

This representation of the density matrix has some advantages over the standard parametrizations. One of them is the possibility to write the state in terms of expectation values of observables (which can be done by expressing the operators $\op{J}_\pm$ in terms of $\op{J}_x$ and $\op{J}_-$). Another advantage is in the study of the entanglement between different systems. The parametrization obtained in this work, allowed us to study the entanglement between subsystems in a bipartite set-up. Interesting examples given were the evaluation of the CHSH inequality for a two-qubit system and for a qubit-qutrit system. In both cases values for the saturation of the non-classical correlations, described by Cirel'son bound, were obtained and briefly discussed.

Since pair-wise entanglement of quantum systems promises to be of benefit to secure quantum information transmission and to quantum computing, and this entanglement is measured via the linear and von Neumann entropies calculated through reduced density matrices, our parametrization allows us to study not only the phase diagrams of interacting systems, but the quantum correlations between subsystems in a bipartite set-up.

Applications of the angular momentum representation for bipartite composite systems, or for even more that two subsystems, can have relevance in the study of entanglement measures, purification procedures, and in general quantum information and quantum computation protocols.

\section*{Acknowledgments}

This work was partially supported by DGAPA-UNAM (under project IN100323). J.A. L.-S. wants to acknowledge the support from the Priority 2030 program at the National University of Science and Technology “MISIS” under the project K1-2022-027.

\section*{Appendix A}
\label{appendixb}

Some other possible parametrizations of the density matrix can be obtained by writing the projectors of the system Eq.~(\ref{projectors}) in a different way. For example, one can have
\begin{equation}
\vert k \rangle \langle l \vert = \frac{1}{(2j)!}\sqrt{\frac{(2j-k+1)!(l-1)!}{(2j-l+1)! (k-1)!}}\op{J}_-^{k-1} \vert j,j \rangle \langle j,-j \vert \op{J}_-^{2j-l+1},      
\end{equation}
where the projector $\vert j,j \rangle\langle j,-j \vert$ can be written in operator form from Eq.~(\ref{ladder}), giving
\begin{equation}
\vert j,j \rangle\langle j,-j \vert=\left(
\begin{array}{cccc}
0 & \cdots & 0 & 1 \\
0 & \ddots & 0 & 0 \\
\vdots & \vdots & \ddots & \vdots \\
0 & \cdots & \cdots & 0
\end{array}
\right)=\frac{\op{J}_+^{2j}}{(2j)!}, 
\end{equation}
which allows us to finally write
\begin{equation}
\vert k \rangle\langle l \vert=\frac{1}{((2j)!)^2} \sqrt{\frac{(2j-k+1)!(l-1)!}{(2j-l+1)!(k-1)!}} \, \op{J}_-^{k-1} \op{J}_+^{2j} \op{J}_-^{2j-l+1}.
\end{equation}

On the other hand one can express the projectors of the system as
\begin{equation}
\vert k \rangle\langle l \vert=\frac{1}{(2j)!}\sqrt{\frac{(k-1)!(l-1)!}{(2j-k+1)!(2j-l+1)!}} \, \op{J}_+^{2j-k+1} \vert j,-j \rangle \langle j,-j \vert \op{J}_-^{2j-l+1}.
\end{equation}
where the projector $\vert j, -j \rangle\langle j,-j \vert$ can be identified by using Eq.~(\ref{ladder}) for $r=s=2j$, resulting in the following angular momentum operator
\begin{equation}
\vert j, -j \rangle\langle j,-j \vert= \left(
\begin{array}{cccc}
0 & \cdots & 0 & 0 \\
\vdots & \ddots & \vdots & \vdots \\
0 & \cdots & \vdots & 0 \\
0 & \cdots & 0 & 1
\end{array}
\right) = \frac{1}{((2j)!)^2} \op{J}_-^{2j} \op{J}_+^{2j} ,
\end{equation}
and in this case the projectors can be obtained as
\begin{equation}
\vert k \rangle\langle l \vert =  \frac{1}{((2j)!)^3}\sqrt{\frac{(k-1)!(l-1)!}{(2j-k+1)!\, (2j-l+1)!}} \, \op{J}_+^{2j -k+1}\op{J}_-^{2j} \op{J}_+^{2j} \op{J}_-^{2j-l+1}.
\end{equation}

\section*{Appendix B}
\label{apC}

In terms of the functional operators
\begin{eqnarray}
\op{F}(\op{J}_z) &:=& \op{J}^2 - \op{J}_z^2 + \op{J}_z= \op{J}_{+}\op{J}_{-}\ ,  \nonumber\\
\op{G}(\op{J}_z) &:=& \op{J}^2 - \op{J}_z^2 - \op{J}_z = \op{J}_{-}\op{J}_{+}\ ,
\end{eqnarray}
and, furthermore, writing $\op{F}(\op{J}_z - k\op{I}) = \op{F}_k$ and $\op{G}(\op{J}_z + k\op{I}) = \op{G}_k$, the ladder operators and their powers may be written as shown in Table~\ref{t1}.
%
\begin{table}[h]
   \caption{Ladder angular momentum operators and their powers in terms of the functional operators $\op{F}_k$ and $\op{G}_k$. As it stands, the entries in table should be read as (row $\times$ column).}
   \begin{center}
   \begin{tabular}{c||c|c|c}
      & $\op{J}_{-}$ & $\op{J}_{-}^2$ & $\op{J}_{-}^3$
      \\ 
      \hline\\[-5mm]
      \hline
      \rule[-4mm]{0mm}{11mm}
      $\op{J}_{+}$ & $\op{F}_0$ & $\op{F}_0 \op{J}_{-}$ & $\op{F}_0\op{J}_{-}^2$
      \\
      \hline
      \rule[-4mm]{0mm}{11mm}
       $\op{J}_{+}^2$ & $\op{F}_1\op{J}_{+}$ & $\op{F}_1 \op{F}_0$ & $\op{F}_1\op{F}_0 \op{J}_{-}$
      \\
      \hline
      \rule[-4mm]{0mm}{11mm}
       $\op{J}_{+}^3$ & $\op{F}_2\op{J}_{+}^2$ & $\op{F}_2 \op{F}_1\op{J}_{+}$ & $\op{F}_2\op{F}_1 \op{F}_0$
   \end{tabular}
   \qquad
    \begin{tabular}{c||c|c|c}
      & $\op{J}_{+}$ & $\op{J}_{+}^2$ & $\op{J}_{+}^3$
      \\ 
      \hline\\[-5mm]
      \hline
      \rule[-4mm]{0mm}{11mm}
      $\op{J}_{-}$ & $\op{G}_0$ & $\op{G}_0 \op{J}_{+}$ & $\op{G}_0\op{J}_{+}^2$
      \\
      \hline
      \rule[-4mm]{0mm}{11mm}
       $\op{J}_{-}^2$ & $\op{G}_1\op{J}_{-}$ & $\op{G}_1 \op{G}_0$ & $\op{G}_1\op{G}_0 \op{J}_{+}$
      \\
      \hline
      \rule[-4mm]{0mm}{11mm}
       $\op{J}_{-}^3$ & $\op{G}_2\op{J}_{-}^2$ & $\op{G}_2 \op{G}_1\op{J}_{-}$ & $\op{G}_2\op{G}_1 \op{G}_0$
   \end{tabular}

   \end{center}
\label{t1}
\end{table}

Using
\begin{equation}
\op{J}_{+}^m \op{J}_{-}^m = \prod_{s=1}^m \op{F}_{m-s} \,, \quad
\op{J}_{-}^m \op{J}_{+}^m = \prod_{s=1}^m \op{G}_{m-s}
\end{equation} 
we can write the general expression for any combination of powers as follows:
\begin{eqnarray*}
m\geq k\,: \quad && \op{J}_{+}^k \op{J}_{-}^m =  \prod_{s=1}^k \op{F}_{k-s}\op{J}_{-}^{m-k}\\
m<k\,: \quad && \op{J}_{+}^k \op{J}_{-}^m =  \prod_{s=1}^m \op{F}_{k-s}\op{J}_{+}^{k-m}
\end{eqnarray*} 
and
\begin{eqnarray*}
m\geq k\,: \quad && \op{J}_{-}^k \op{J}_{+}^m =  \prod_{s=1}^k \op{G}_{k-s}\op{J}_{+}^{m-k}\\
m<k\,: \quad && \op{J}_{-}^k \op{J}_{+}^m =  \prod_{s=1}^m \op{G}_{k-s}\op{J}_{-}^{k-m} \,.
\end{eqnarray*} 

As an example, the density matrix for the ququart in terms of these functional operators takes the form
\begin{equation}
\op{\rho} = \left(\begin{array}{cccc}
\langle \op{F}_2 \op{F}_1 \op{F}_0 \rangle &  \langle \op{F}_1 \op{F}_0 \op{J}_{-} \rangle & \langle \op{F}_0 \op{J}_{-}^2 \rangle & \langle \op{J}_{-}^3\rangle\\[3mm]
\langle \op{F}_2 \op{F}_1 \op{J}_{+} \rangle & \langle \op{F}_1 \op{F}_0 \op{G}_0 \rangle & \langle \op{F}_0 \op{G}_1 \op{J}_{-} \rangle & \langle \op{G}_2 \op{J}_{-}^2 \rangle \\[1mm]
\langle \op{F}_2 \op{J}_{+}^2 \rangle & \langle \op{G}_0 \op{F}_1 \op{J}_{+}\rangle & \langle \op{F}_0 \op{G}_1 \op{G}_0 \rangle & \langle \op{G}_2 \op{G}_1 \op{J}_{-}\rangle \\[1mm]
\langle \op{J}_{+}^3 \rangle & \langle \op{G}_0 \op{J}_{+}^2 \rangle & \langle \op{G}_1 \op{G}_0 \op{J}_{+} & \langle \op{G}_2 \op{G}_1 \op{G}_0 \rangle
\end{array} \right)\, .
\end{equation}

\section*{References}



\providecommand{\newblock}{}

\end{document}